\documentclass[12pt]{article}
\usepackage{cite}
\usepackage{epsfig}
\usepackage{graphicx}

\title{\LARGE\bf{Field emission properties of nano-composite carbon nitride films}\\}

\author{\large\bf {I. Alexandrou$^{1}$, M. Baxendale$^2$,
G. A. J. Amaratunga$^2$}\\ {\large\bf {N. L. Rupesinghe$^2$ and C.
J. Kiely$^1$}}}

\begin{document}
\date{ }
\maketitle
\begin{center}
\small {$^1$ Department of Engineering, Materials Science $\&$
Engineering, University of Liverpool, Liverpool L69 3BX, UK.\\
 $^2$ Department of Engineering, University of Cambridge,
 Cambridge CB2 1PZ, UK}.
 \vspace{0.5cm}
\end{center}

\renewcommand{\baselinestretch}{1.5}
\small\normalsize
\thispagestyle{empty}
\begin{abstract}
A modified cathodic arc technique has been used to deposit carbon
nitride thin films directly on n$^+$ Si substrates. Transmission
Electron Microscopy showed that clusters of fullerene-like
nanoparticles are embedded in the deposited material. Field
emission in vacuum from as-grown films starts at an electric field
strength of 3.8~V/$\mu$m. When the films were etched in an
HF:NH$_4$F solution for ten minutes, the threshold field decreased
to 2.6~V/$\mu$m. The role of the carbon nanoparticles in the field
emission process and the influence of the chemical etching
treatment are discussed.
\end{abstract}
%\newpage
%
\section{Introduction}
~~~ One of the applications of carbon based materials studied over
the last few years has been the production of flat cold cathodes
which can serve as field emitters with low turn-on voltages and
high emission currents. The first carbon based films that
triggered attention in terms of their field emission properties
were a-C:H, a-C:N and diamond-like (DLC)
coatings~\cite{400,401,394,339,337,336,340,338}. However, after
the discovery of carbon nanotubes and the development of
techniques for their production in macroscopic quantities, there
has been an increasing interest in their field emission
characteristics~\cite{333,334,335}. In such studies, single walled 
and multi-walled carbon nanotubes were derived from the carbon
soot produced by an electric arc discharge in high helium (or
other inert background) gas pressure. The carbon nanotubes were
then separated from the rest of the soot and were either deposited on an appropriate %%@
flat surface in a suspension or, alternatively,macroscopic bundles of tubes were studied %%@
directly. The field emission characteristics of such structures are generally much %%@
better than that of the a-C:N and DLC films, but their use in practical applications is %%@
very difficult because of the inability to directly form a thin film. Carbon nanotubes %%@
display a metallic behaviour~\cite{379} with states very near to the Fermi %%@
level~\cite{249}. The ionisation potentials and electron affinities are 6.6~eV and %%@
4.3~eV for open tubes and 6.7~eV and 3.6~eV for capped tubes, respectively. These values %%@
have been calculated by Lou {\it {et al}}~\cite{206} using a density-functional cluster %%@
method for nanotubes consisting of 204 carbon atoms. The effect of an externally applied %%@
electric field along the tube axis (towards the tube base) has also been %%@
studied~\cite{206}. It was shown that as the electric field
increases, the electrons became redistributed in such a way as to over populate the tip %%@
of the tube, which contributes to an increased concentration of the external electric %%@
field around the tip region. Although the ionisation potential and the electron affinity %%@
of the nanotubes are higher than for other materials, the field enhancement due to %%@
geometrical factors (i.e. the high aspect
ratio) is expected to decrease the magnitude of the threshold
external electric field $\mathcal {E}_{th}$ required to commence
field emission. Also, the electron transport properties described
by Lou {\it {et al}}~\cite{206} predict that once field emission
starts, high emission current values can be realised and
maintained. These properties make carbon nanotubes very attractive
candidates as field emitters. Indeed, previous work~\cite{334} on
individual carbon nanotubes has revealed currents of 1~$\mu$A for
a field strength of 2.2~V/$\mu$m. Furthermore, for a nanotube
suspension film, a current density of 1~$\mu$A/cm$^2$ was achieved
for a field strength of 3.3~V/$\mu$m.

In this work, as-grown and chemically etched nano-composite CN$_x$
films are tested for their field emission properties, while High
Resolution Electron Microscopy (HREM) is used to study their
microstructure. The field emission characteristics of the
nano-composite film are compared to those of a nanotube suspension
material. The role of structure and chemical etching on the field
emission properties of the deposited material are also discussed.
\section{Experimental details}
\begin{figure}[tbp]
\begin{center}
\renewcommand{\baselinestretch}{1.2}
\rotatebox{270}{\resizebox{!}{11cm}{\includegraphics{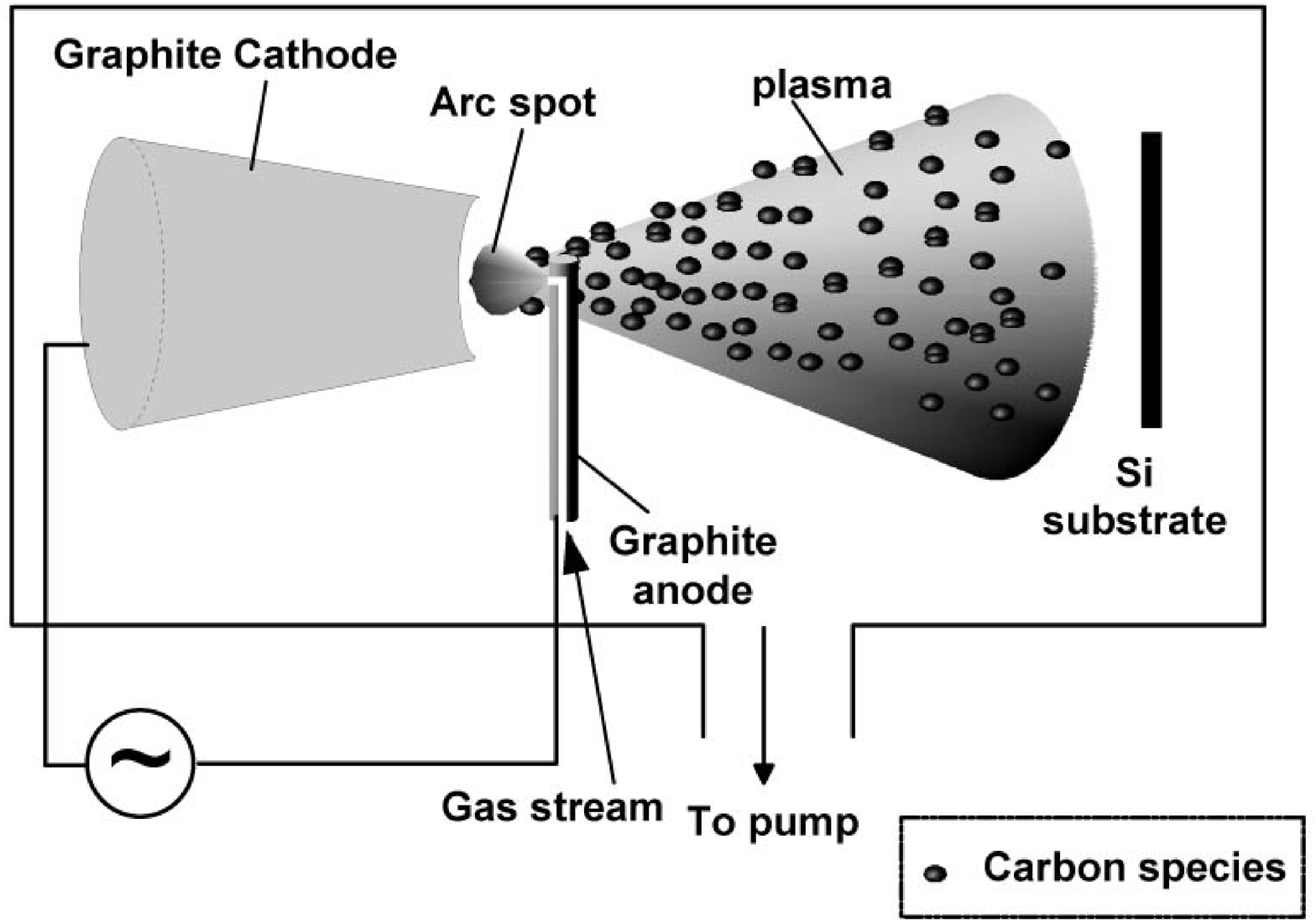}}}
\caption{\small Schematic representation  of the anodic jet carbon
arc deposition technique.}
\label{anode-jet}
\renewcommand{\baselinestretch}{1.5}
\small\normalsize
\end{center}
\end{figure}

~~~~ Figure~\ref{anode-jet} shows a  schematic representation of
the experimental configuration of the Anodic Jet Carbon Arc (AJCA)
deposition technique. N$_2$ gas was injected into the chamber and
was directed towards the arcing region through an orifice at the
side of a specially machined hollow anode. The background pressure
in the deposition chamber was used to control the flow of
nitrogen. The ionisation of nitrogen during the discharge was
indicated by the characteristic intense pink glow. The working
pressure in the chamber was kept below 1$\times$10$^{-3}$~mbar.
For all operating conditions, a drop of 2$\times$10$^{-4}$~mbar in
the nitrogen background pressure was noted during the arcing
period. The substrate was positioned  27~cm away from the conical
electrode. The voltage applied between the cathode and anode was
the $a.c.$ output of a power supply used for arc welding, which
had an output voltage of 22-24~V $a.c.$, a frequency of 50~Hz and
could deliver a current between 57 and 140~A. The use of an $a.c$
voltage is not very common and effectively results in each
electrode oscillating between being an anode and a cathode. The
electric arc was therefore interrupted 100 times every sec (i.e. every time the $a.c$ %%@
power became zero), creating an effective pulse rate of 100~Hz.

Using this configuration, films were  deposited simultaneously
onto unheated freshly cleaved NaCl $\{100\}$ crystals and onto n$^+-$type Si wafers. In %%@
this way, specimens for electron microscopy characterisation could be easily prepared by %%@
dissolving the NaCl crystal in distilled water, thus avoiding the use of strong acid %%@
solutions including HF, whereas the samples grown on n$^+-$type Si wafers were used in %%@
the field emission studies.

Two types of nanotube were used to form the suspension films; multi-walled nanotubes %%@
formed using the arc method~\cite{382} and bundles of single walled nanotubes obtained %%@
using a catalytic
method based on creating a carbon arc between electrodes loaded with Ni or Y~\cite{402}. %%@
The latter type has a catalytic metal particle situated at the tip of the tube. The %%@
suspension films
were formed by filtering a nanotube suspension in hexane through a
microporous (0.2~$\mu$m) paper. The final nanotube film was the
dried residue on the filter paper, which appeared as a uniform grey deposit. Electrical %%@
contact between the film and the substrate was acquired through silver dag at the edges %%@
of the substrate.

The field emission measurements from the arc deposited films and from the nanotube %%@
samples were performed in a vacuum of 10$^{-8}$~mbar by using the examined samples as %%@
cathodes and collecting the emitted electrons on an Al anode with an effective %%@
collection area of 0.5~cm$^2$. The spacing between the anode and the cathode was chosen %%@
to be between 80 and 100~$\mu$m and was defined by separating the electrodes using %%@
optical grade glass fibers with well specified diameters. The voltage between the anode %%@
and the cathode was varied while simultaneously recording the current using combined %%@
source/measure electrometers.
\section{Results}
\subsection{Field emission measurements}
\begin{figure}[t]
\begin{center}
\renewcommand{\baselinestretch}{1.2}
\resizebox{!}{17.0cm}{\includegraphics{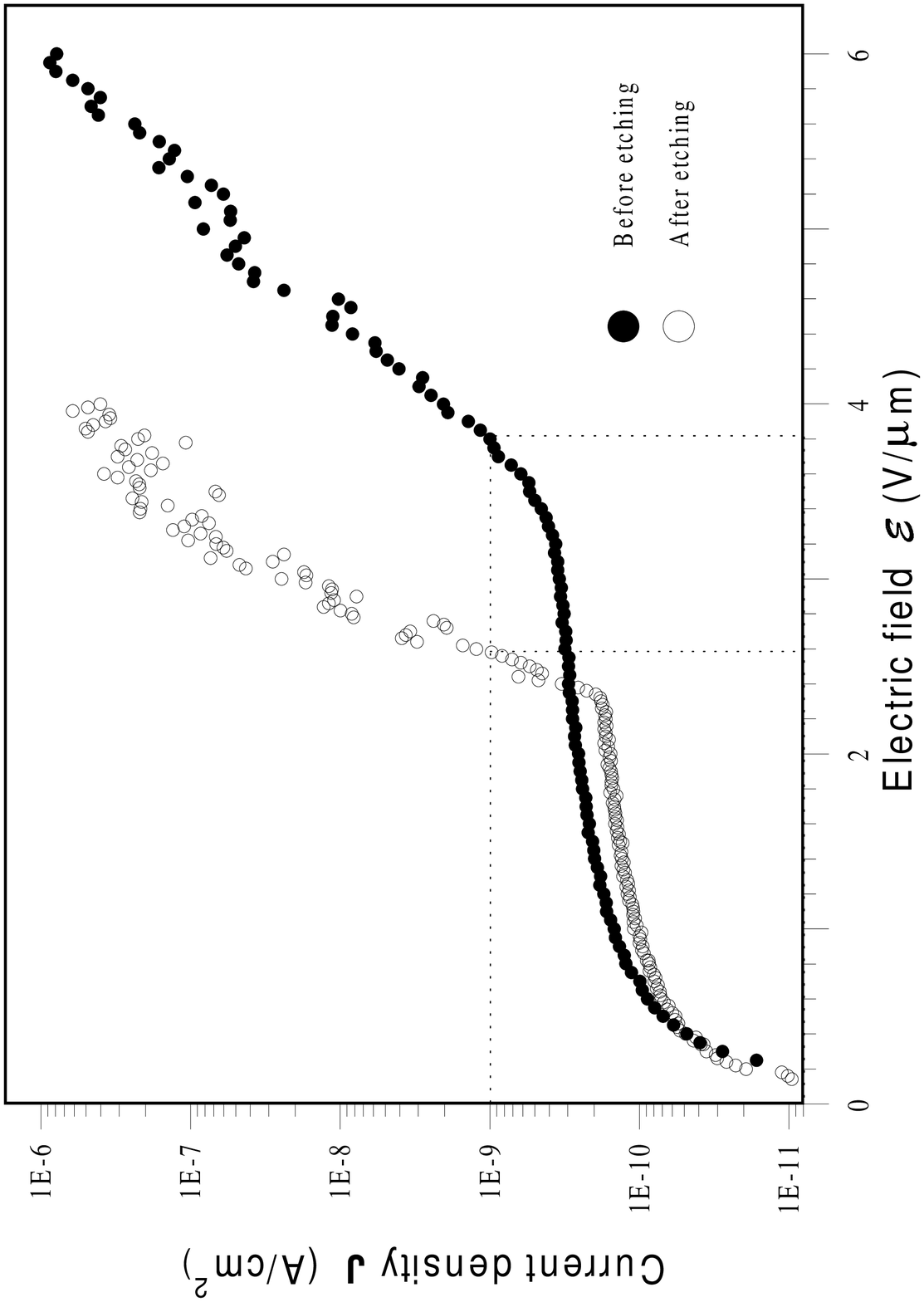}} \caption{\small
Current density vs. applied electric field for the same
material before and after it was etched for 10 minutes using an
HF:NH$_4$F solution. The particular material was
deposited at room temperature, using an arc current of 140~A with
a nitrogen jet creating a background pressure of 1$\times
10^{-3}$~mbar.} \label{j_e}
\renewcommand{\baselinestretch}{1.5}
\small\normalsize
\end{center}
\end{figure}

\begin{figure}[t]
\begin{center}
\renewcommand{\baselinestretch}{1.2}
\resizebox{!}{17.0cm}{\includegraphics{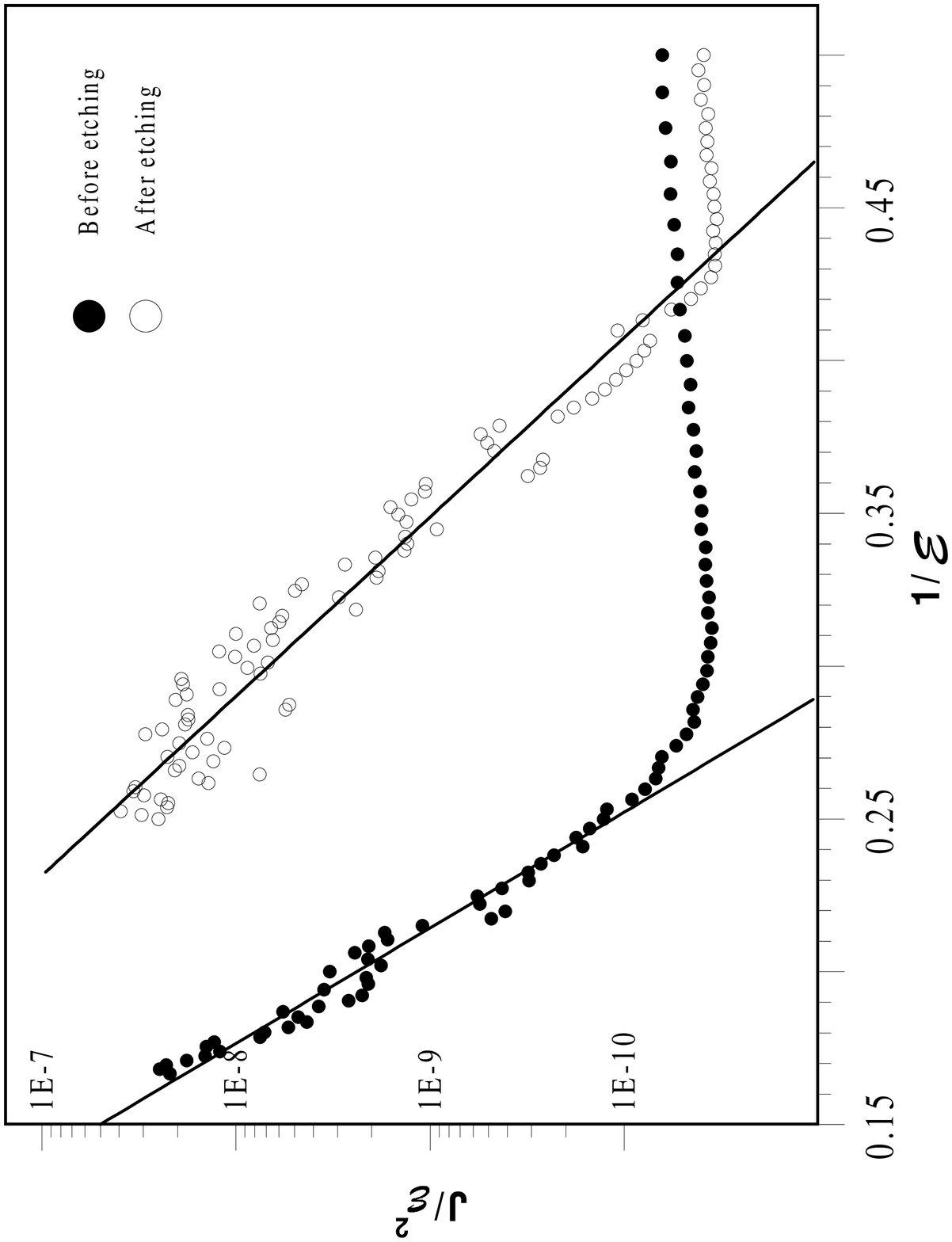}}
 \caption{\small The Fowler-Nordheim plot of the
experimental data shown in Fig.~\ref{j_e}. A mathematical fit of
the form of $J/\mathcal {E}^2 = A$ exp$(B/\mathcal {E})$ on the
experimental data is a straight line.} \label{f_n}
\renewcommand{\baselinestretch}{1.5}
\small\normalsize
\end{center}
\end{figure}

~~~~ As-grown films deposited on n$^+$-Si substrates using the AJCA method were tested %%@
for their field emission properties. The effect of chemical etching on the field %%@
emission characteristics was also studied by dipping the samples in an HF:NH$_4$F %%@
solution for various times and then re-measuring the field emission properties. The %%@
field emission current density $J$~(A cm$^{-2}$) has been calculated by dividing the %%@
measured current by the emission area. Figure~\ref{j_e} shows representative $J-\mathcal %%@
{E}$ curves obtained during field emission measurements before and after chemical %%@
etching in an HF:NH$_4$F solution, where $\mathcal {E}$ is the applied electric field. %%@
In this work, the threshold field ($\mathcal {E}_{th}$) is defined as that required to %%@
cause a field emission current density of 10$^{-9}$~A/cm$^2$. For the as-grown film, %%@
$\mathcal {E}_{th}$ was 3.8~V/$\mu$m which decreased to 2.6~V/$\mu$m after the film had %%@
been chemically etched for ten minutes. Chemical etching for longer periods resulted in %%@
film delamination because the Si substrate started to dissolve. Furthermore, etching for %%@
periods shorter than 10 minutes resulted in a smaller decrease in $\mathcal {E}_{th}$. %%@
The respective Fowler-Nordheim (F-N) plots are shown in Fig.~\ref{f_n}. In these %%@
diagrams $\log_{10} J/\mathcal {E}^2$ is plotted vs. 1/$\mathcal {E}$. A mathematical %%@
fit on the experimental data of the form $J/\mathcal {E}^2 = A \exp[B/\mathcal {E}]$, is %%@
a straight line (in both cases), revealing that the electron current density follows the %%@
F-N equation: 
\begin{eqnarray}
J=6.2\times 10^{-2}\frac{\sqrt{f / \phi}}{f + \phi}\mathcal{E}^2
\exp\left( -6.8\times10^{3}\frac{\phi^{3/2}} {\gamma \mathcal
{E}}\right) \label{F_N}
\end{eqnarray}
which means that the observed current comes from field emission. In %%@
equation~(\ref{F_N}), $\mathcal {E}$ is in V/$\mu$m whereas $\phi$ and $f$ represent (in %%@
eV) the work function and the energy difference between the top of the valence band and %%@
the Fermi level, respectively. $\gamma$ is the field enhancement factor and is used to %%@
denote the field concentration mainly due to topographical characteristics. By assuming %%@
that $\phi$ and $f$ do not change with etching, the field enhancement factor, $\gamma$, %%@
is found to have increased by a factor of 1.55 as a result of etching the film.

\begin{figure}[tbp]
\begin{center}
\renewcommand{\baselinestretch}{1.2}
\rotatebox{0}{\resizebox{!}{20.0cm}{\includegraphics{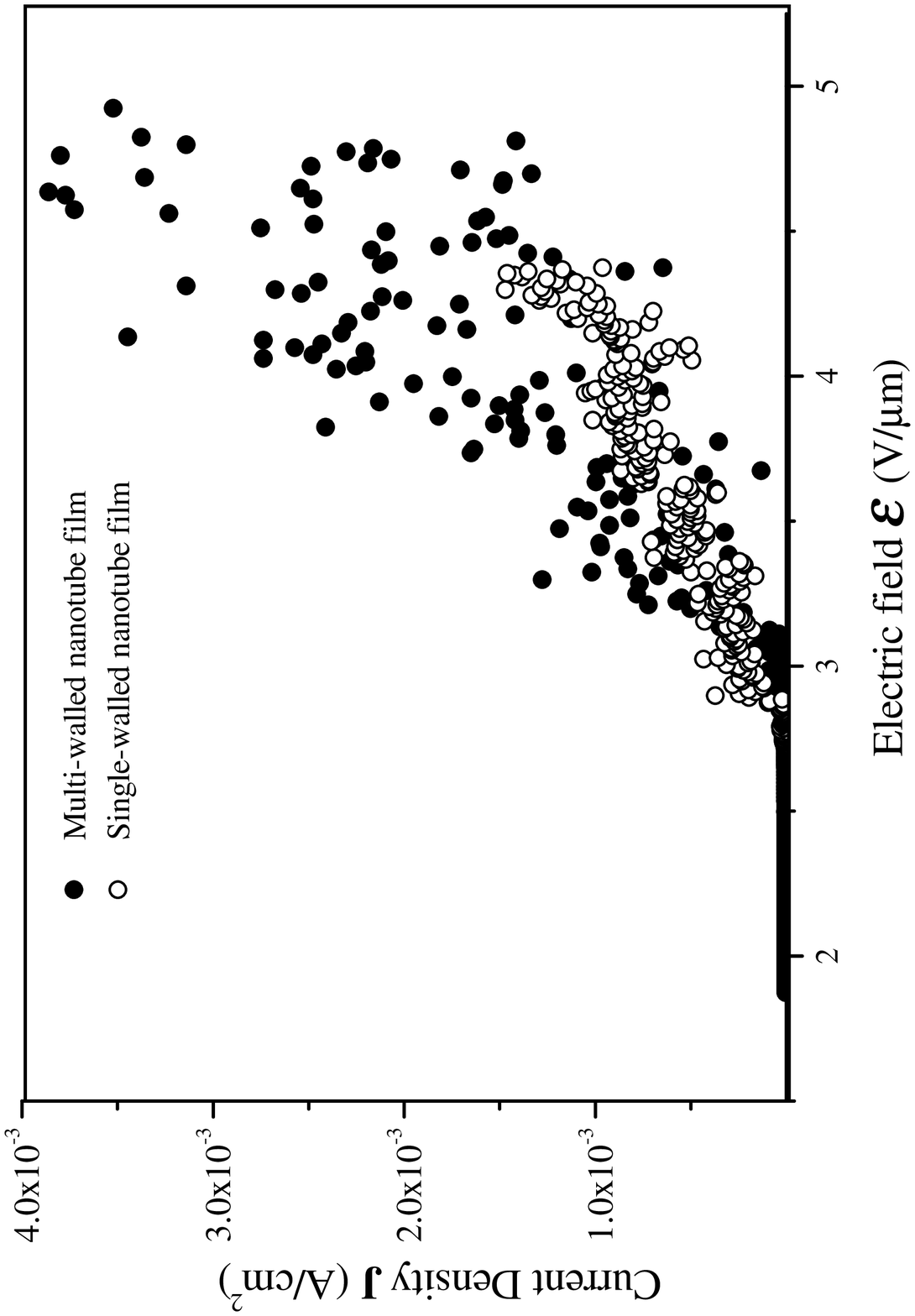}}}
\caption{\small Current density vs. applied electric field for
films of pure single-walled and multi-walled carbon nanotubes. The
anode to cathode spacing was 100~$\mu$m.} \label{j_e_nt}
\vspace{5cm}
\renewcommand{\baselinestretch}{1.5}
\small\normalsize
\end{center}
\end{figure}

\begin{figure}[tbp]
%\vspace{15cm}\begin{center}
\renewcommand{\baselinestretch}{1.2}
\begin{center}
\rotatebox{0}{\resizebox{!}{20.0cm}{\includegraphics{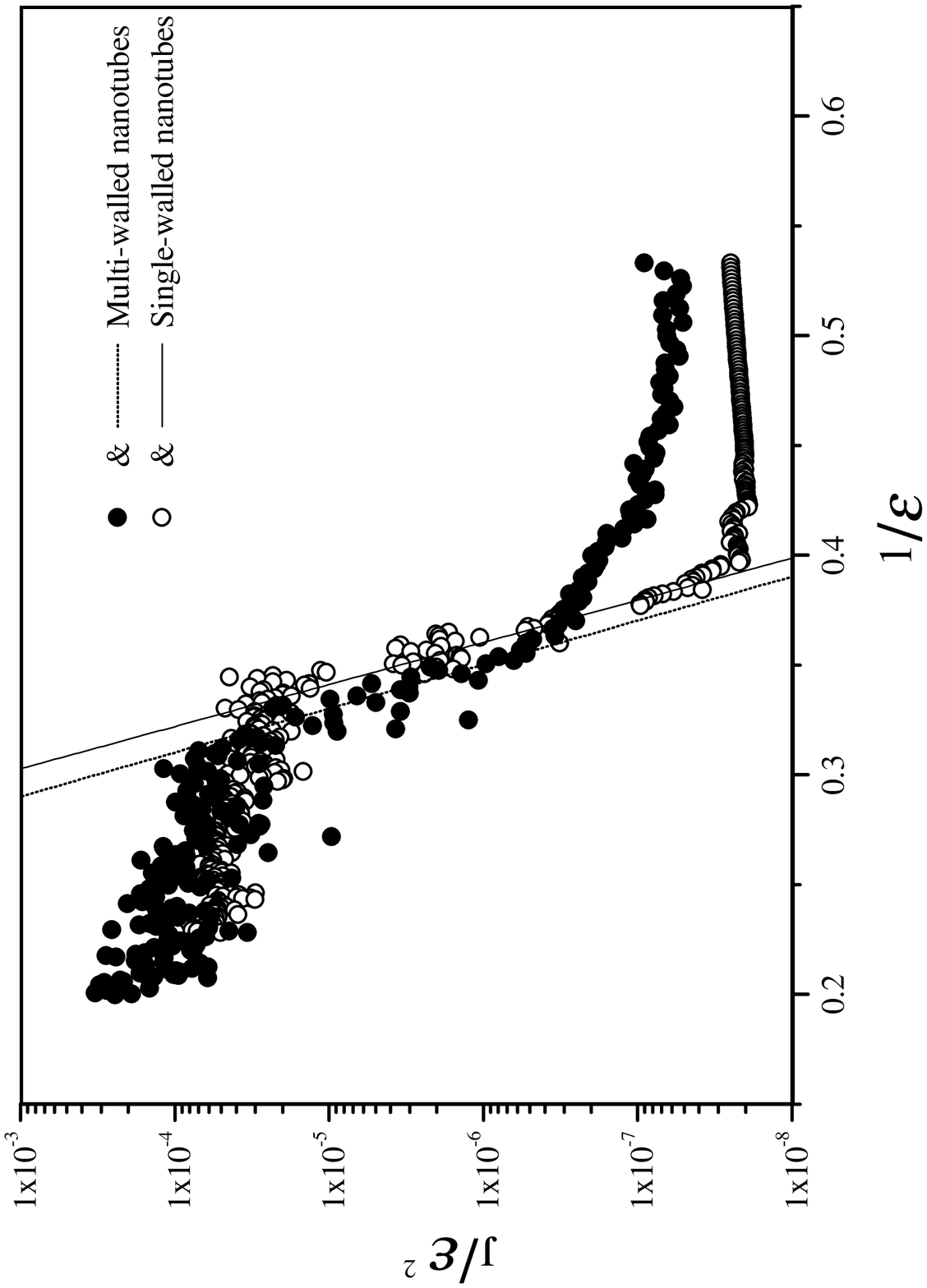}}}
\caption{\small The Fowler-Nordheim plot of the experimental data
shown in Fig.~\ref{j_e_nt}. The experimental data that correspond
to field emission are fitted with an equation of the form of
$J/\mathcal {E}^2 = A$ exp$(B/\mathcal {E})$.} \label{f_n_nt}
\renewcommand{\baselinestretch}{1.5}
\small\normalsize
\end{center}
\end{figure}

Results from field emission measurements on randomly oriented pure carbon nanotube films %%@
formed by suspension evaporation are shown in Fig.~\ref{j_e_nt}. The respective %%@
Fowler-Nordheim (F-N) plots are shown in Fig.~\ref{f_n_nt}. As is expected from %%@
consideration of the metallic nature of the carbon nanotubes, there is a well defined %%@
F-N type threshold field, after which there is an exponential current rise region, %%@
followed by saturation. Interestingly, the arc deposited and the pure nanotube films %%@
have similar field emission behaviour. However, during the field emission measurements %%@
of the pure nanotube films the background current density is higher than 1~$\mu$A/cm$^2$ %%@
so the value of the threshold field cannot be defined as above. Instead, by considering %%@
Fig.~\ref{j_e_nt}, a clear increase is seen in the emitted current density at a field of %%@
about 2.85~V/$\mu$m for the single-walled nanotube film and 3.2~V/$\mu$m for the %%@
multi-walled nanotube film. The similar field emission behaviour between the arc %%@
deposited and the nanotube films suggests that both materials include the same emission %%@
sites, but their number density being considerably higher in the ``suspension" nanotube %%@
films.

From the F-N plot of the carbon nanotube emitters, field enhancement factors, $\gamma$, %%@
of 661 and 633 are calculated for the multi-walled and the single-walled nanotubes, %%@
respectively, given that the work function, $\phi$ has a value of 5~eV~\cite{334,379}. %%@
If the same value of $\phi$ is used for the emitting sites of the nano-composite film, %%@
$\gamma$ is 1432 for the as-grown films and 2236 for the chemically etched film.
\subsection{High resolution electron microscopy}
\begin{figure}[tbp]
\begin{center}
\rotatebox{0}{\resizebox{!}{20.0cm}{\includegraphics{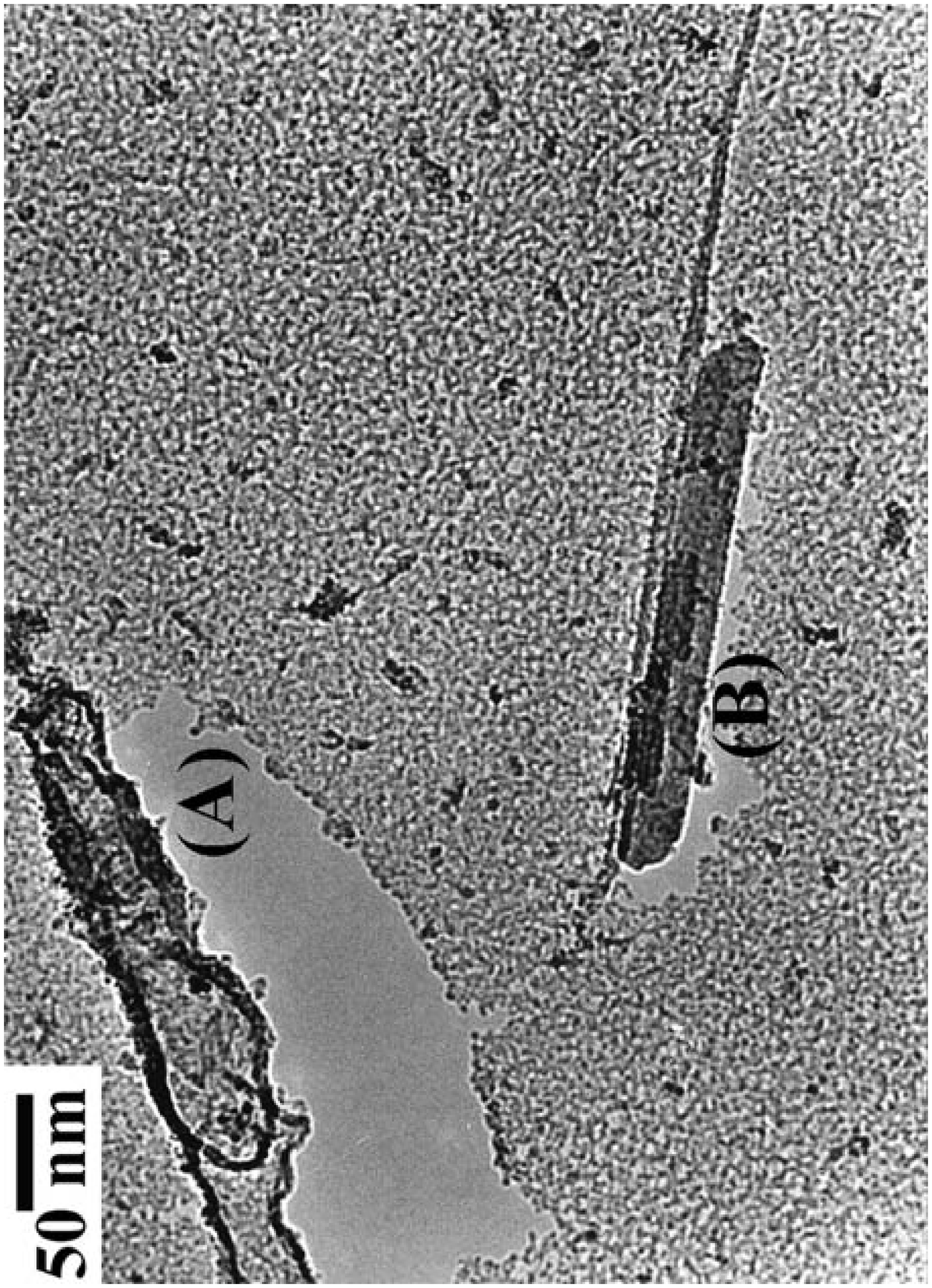}}}
\end{center}
\renewcommand{\baselinestretch}{1.2}
\caption{\small Low magnification TEM image of  a film deposited
using AJCA with an N$_2$ jet creating a pressure of
0.8$\times$10$^{-3}$~mbar.} \label{ncac_tem}
\renewcommand{\baselinestretch}{1.5}
\small\normalsize
\end{figure}

\begin{figure}[tbp]
\begin{center}
\rotatebox{0}{\resizebox{!}{20.0cm}{\includegraphics{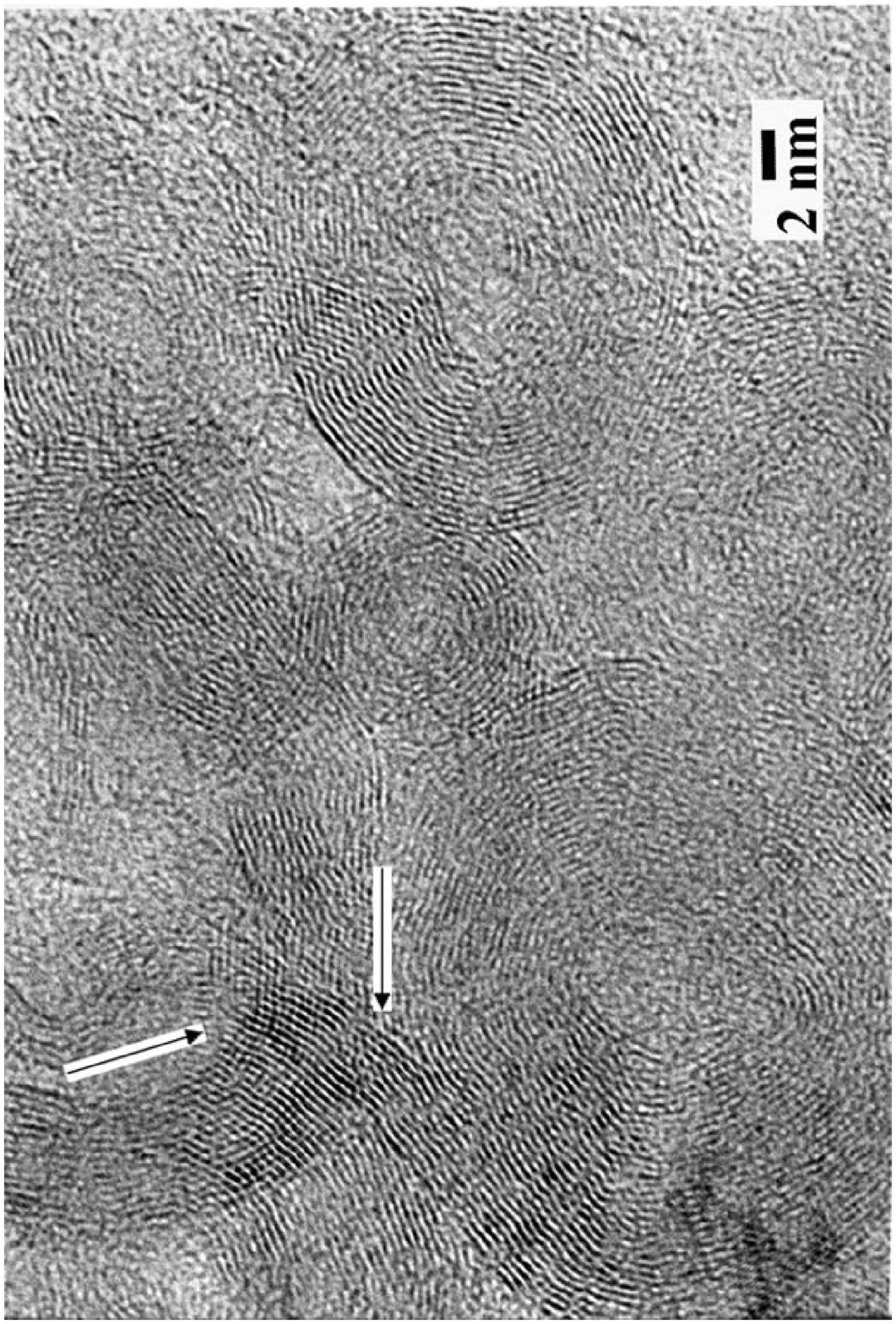}}}
\end{center}
\renewcommand{\baselinestretch}{1.2}
\caption{\small HREM image of a typical cluster of nanoparticles
in a film deposited using AJCA.} \label{ncac_tem2}
\renewcommand{\baselinestretch}{1.5}
\small\normalsize
\end{figure}

~~~~ Electron transparent specimens for TEM investigation were
prepared by dissolving away the NaCl substrate in distilled water
or in the HF:NH$_4$F solution and the free-standing films were
then floated onto Cu mesh grids. By comparing the two different
samples we could investigate the effect of chemical etching on the
film microstructure. Figure~\ref{ncac_tem} shows a low
magnification TEM image of a film deposited using the AJCA technique in a background %%@
pressure of 0.8$\times$10$^{-3}$~mbar. The electron beam has been strongly underfocussed %%@
(i.e. much more than Scherzer defocus) in order to obtain stronger contrast. In such %%@
conditions, the diffracting species appear darker than the background. Two kinds of %%@
diffracting features can be seen in this image. The two very large diffracting areas %%@
(labelled {\bf A} and {\bf B}) are $\mu$m scale macroparticles containing long graphitic %%@
layers reminiscent of turbostratic carbon. Secondly, there are numerous small %%@
diffracting areas (a few of which are labelled C) dispersed almost uniformly throughout %%@
the film, their size varying between 10 and 20~nm. Observation at higher magnification %%@
under phase contrast imaging conditions (such as in Figure~\ref{ncac_tem2}) revealed %%@
that most of the latter areas are clusters of fullerene-like nanoparticles. Sets of %%@
concentric graphene sheets with interlayer spacing of 3.4~\AA~are clearly seen in %%@
Fig.~\ref{ncac_tem2}. The pattern formed resembles closely packed multi-walled carbon %%@
nanotubes imaged along their axes (i.e. aligned along the direction of film growth). %%@
Alternatively, the particles seen in Fig.~\ref{ncac_tem2} could also be more spherical %%@
bucky ``onion"
type entities. Some of them are distorted with several graphene
planes buckling away  (at least in projection) towards other nanoparticles (as indicated %%@
by arrows). It is proposed that the observed nanoparticle agglomerates are included in %%@
the condensing flux~\cite{mine,physrevb}, rather than formed on the film. Therefore, the %%@
agglomerates observed in the sample grown on the NaCl substrate are expected to be %%@
identical to those in the film deposited on n$^+$-Si, used for field emission %%@
measurements. 

\begin{figure}[tbp]
\begin{center}
\renewcommand{\baselinestretch}{1.2}
\rotatebox{0}{\resizebox{!}{15.0cm}{\includegraphics{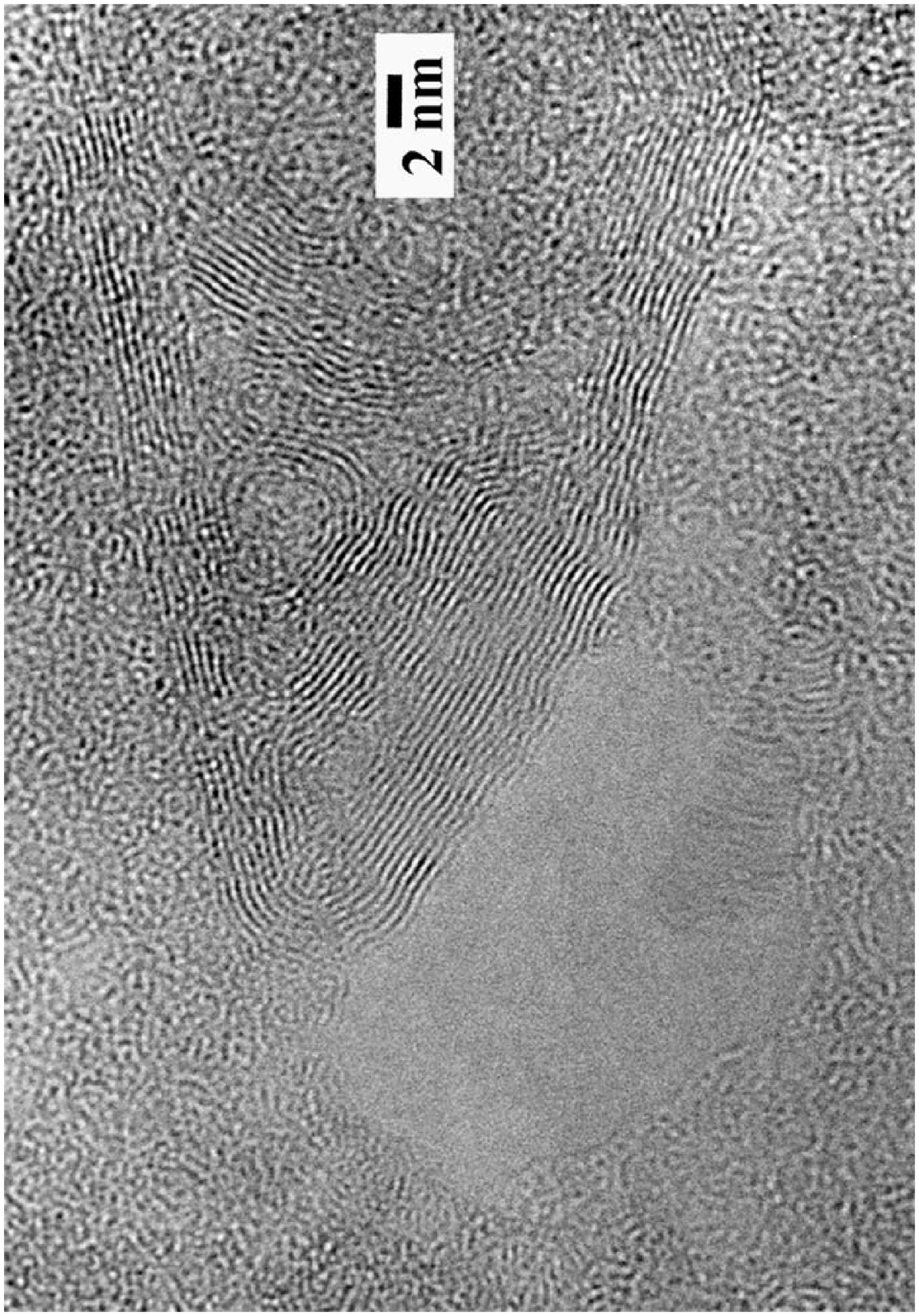}}}
\vspace{0.5cm} \caption{\small Characteristic hole on the film
surface after ten of minutes chemical etching. Graphene sheets can
be clearly seen on one side of the hole.} \label{ncac_etched_tem}
\renewcommand{\baselinestretch}{1.5}
\small\normalsize
\end{center}
\end{figure}

When the same film was examined after it had been etched for 10 minutes the surface %%@
density of large macroparticles such as {\bf {A}} and {\bf {B}} shown in %%@
Fig.~\ref{ncac_tem}, decreased dramatically and pin-holes became apparent in the film %%@
surface. Figure~\ref{ncac_etched_tem} shows a plan-view HREM image from an area thinned %%@
by the chemical etching. In the majority of cases, we could image fullerene-like fringe %%@
structures in the vicinity of the pinhole. The removal of the large graphitic particles %%@
suggests that they were not strongly bonded to the film but simply lying on the film %%@
surface. Also, the preferential thinning of some areas and particularly the existence of %%@
nanoparticles at the edge of the pinholes suggests that the amorphous material %%@
surrounding the embedded nanoparticles was preferentially removed by the HF:NH$_4$F %%@
solution. Similar behaviour has been noted previously during the preparation of TEM %%@
samples of CN$_x$ films deposited by carbon arc techniques~\cite{nature,emag97}, where %%@
an HF:HNO$_3$:H$_2$O solution was used to dissolve away the Si substrate. The %%@
preferential removal of the amorphous material in such a way may be connected to the %%@
better field emission properties obtained after chemical etching. Furthermore, the easy %%@
removal of material surrounding the carbon nanoparticles suggests that they were not %%@
strongly bonded to their surrounding material, but rather they were produced on the %%@
cathode and/or the arcing region and were then subsequently embedded in the growing %%@
film.
\section{Discussion}
~~~~ The nanoparticles included in our films contain de-localised sp$^2$ states that are %%@
expected to promote the transfer of electrons from the bulk of the material to the %%@
surface via the nanotube itself, thus enhancing the field emission current density. %%@
However, it is the nanoparticle shape that is expected to govern field emission %%@
parameters such as the threshold field $\mathcal{E}_{th}$. Given the comparable field %%@
emission characteristics between the nano-composite films and the nanotube suspension %%@
films, the role of the nanoparticles in the nano-composite films is considered here by %%@
correlating the field emission results with HREM observations.

HREM imaging experiments reveal that clusters of fullerene-like nanoparticles occur at a %%@
high density in the AJCA films. When such nanoparticles protrude from the film surface, %%@
they create the correct surface topography for local enhancement of the external %%@
electric field. Nanoparticles having any orientation with respect to the external %%@
electric field will enhance the field emission properties of the material as long as a %%@
fraction of their length intersects the surface of the film. The agglomeration of %%@
fullerenes shown in Fig.~\ref{ncac_tem2} represents a cluster with the ideal orientation %%@
since all nanoparticles seem to be aligned normal to the film surface. When an external %%@
electric field is applied, the semi-metallic nanoparticles would cause a field %%@
concentration around their tips, increasing the electric field strength locally. %%@
Therefore, although the magnitude of the applied external field should not be large %%@
enough to induce field emission from the sp$^2$-bonded carbon layers, its local value in %%@
the regions of highest curvature of the fullerene could be concentrated just enough to %%@
exceed $\mathcal{E}_{th}$.

The field concentration will depend on how much of the nanoparticle protrudes above the %%@
film surface and on the diameter of its tip. Hence the aspect ratio of that part of the %%@
embedded nanoparticle that stands proud from the film surface controls the local value %%@
of $\mathcal{E}_{th}$. The importance of the aspect ratio is evident from embedded %%@
cluster calculations of Lou {\it {et al}}~\cite{206}, which predict that for aspect %%@
ratios of 20 and 100 the field enhancements will be 12 and 80, respectively. These %%@
calculations reveal that the field enhancement increases faster than a linear rate with %%@
respect to the aspect ratio. In addition, when several nanoparticles are grouped %%@
together (as in Fig.~\ref{ncac_tem2}), the field concentration is expected to depend on %%@
the aspect ratio of the group and not on the aspect ratios of the individual %%@
nanoparticles. This is because the electric field applied externally will not change %%@
significantly from tip to tip when the respective nanoparticles are very close together. %%@
Therefore, a cluster will approximate to a single large nanoparticle having the same %%@
dimensions. Thus, the grouping of $k$ identical nanoparticles would decrease both the %%@
aspect ratio and the field enhancement factor. This rather simple model could explain %%@
the difference in the field required (2.2~V/$\mu$m,~\cite{334}) to obtain a current of %%@
1~$\mu$A from an individual carbon nanotube and a current density of 1~$\mu$A/cm$^2$ %%@
from a film containing clusters of carbon nanoparticles (3.3~V/$\mu$m,~\cite{334}, and %%@
$\approx$4.2~V/$\mu$m for our chemically etched films). Similarly, the nano-composite %%@
films appear to have higher field enhancement factors, even though the nanoparticles are %%@
included in the arc material are expected to have a much smaller aspect ratio than %%@
single-walled and multi-walled nanotubes. This can be attributed to the fact that the %%@
nanotubes in the suspension film exist at such high densities that they create a uniform %%@
background of emerging tips. Therefore only a small part of each nanotube is emerging %%@
from the background decreasing the ``effective" aspect ratio and their close proximity %%@
prevents the concentration of the electric field on each of them.

Furthermore, there is a decrease in the threshold field, with a simultaneous increase in %%@
the emission current, when the nano-composite film was etched in the HF:NH$_4$F %%@
solution. Firstly, chemical etching was found to remove the majority of the loosely %%@
bonded macroparticles. The improved field emission characteristics after the chemical %%@
etching treatment implies that the emission does not originate from such sites. This %%@
supports the suggestion that the emission current is from nanoparticles such as those %%@
seen in Fig.~\ref{ncac_tem2}. Secondly, during deposition, we expect that C:N amorphous %%@
material will totally bury the majority of the nanoparticles. Such an overlayer might %%@
locally create an additional potential barrier to electron emission. HREM images such as %%@
that shown in Fig.~\ref{ncac_etched_tem} revealed that after etching the film is thinner %%@
in the locality of the nanoparticles. This suggests that the nanoparticles are weakly %%@
bonded to the matrix material, which allows for the easy removal of the C:N material %%@
adjacent to the nanoparticle. The weak bonding of the larger carbon nanoparticles to the %%@
film matrix is also consistent with the suggestion that they are formed at the cathode %%@
region and are subsequently embedded in the growing film. Consequently, the role of %%@
chemical etching seems mainly to be in cleaning the sharp ends of the nanoparticles, %%@
thus improving their emission properties and secondly, totally ``buried" nanoparticles %%@
become partially uncovered. The latter effect can explain the increase in the emission %%@
current density when the AJCA films were etched.

\begin{figure}[tbp]
\begin{center}
\renewcommand{\baselinestretch}{1.2}
\rotatebox{0}{\resizebox{!}{17.0cm}{\includegraphics{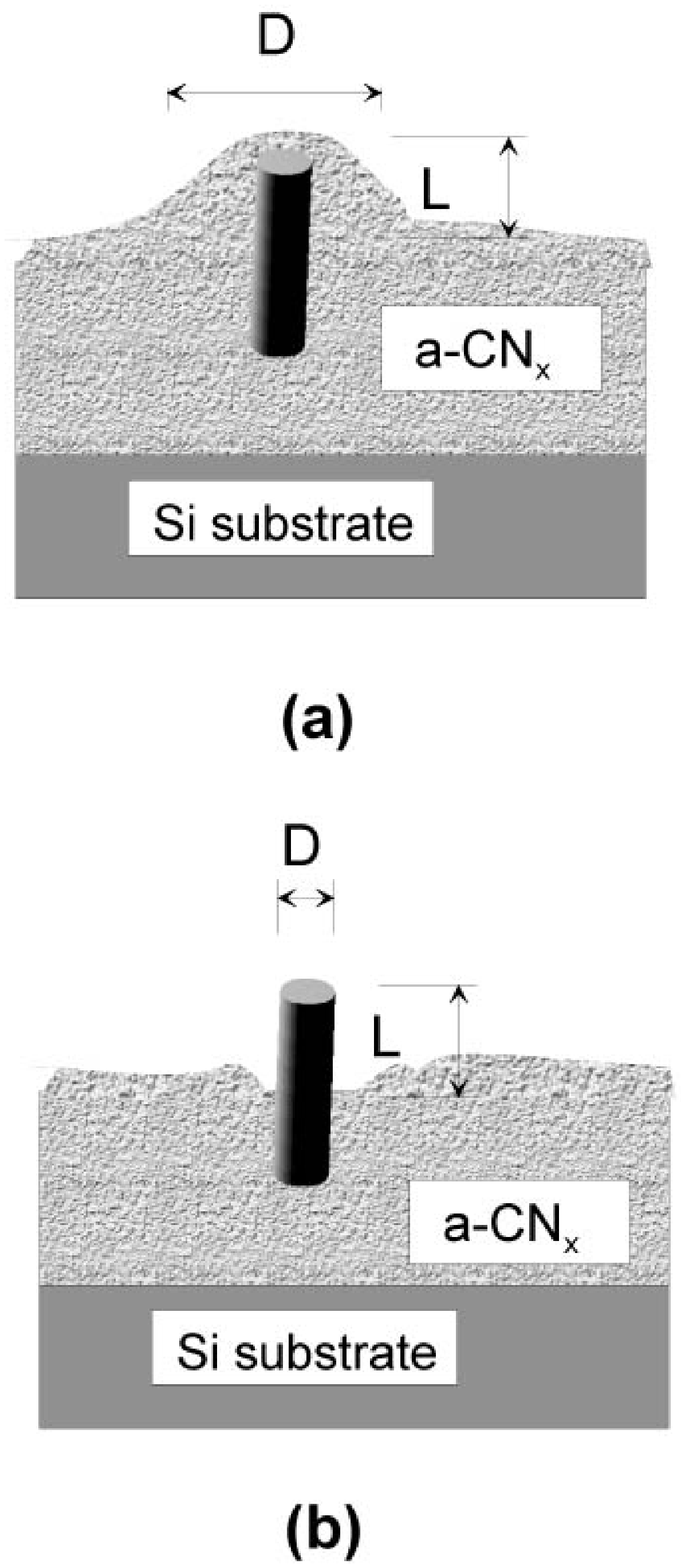}}}
\caption{\small Schematic illustration of the effect of the
chemical etching on the aspect ratio of nanoparticles or their
clusters.} \label{etching}
\renewcommand{\baselinestretch}{1.5}
\small\normalsize
\end{center}
\end{figure}

A possible effect of chemical etching on the aspect ratio of the clusters of carbon %%@
nanoparticles is shown schematically in Fig.~\ref{etching}. Before chemical etching, a %%@
cluster or a single nanoparticle is covered with amorphous material. It proposed that %%@
the main reason for this is that clusters of nanoparticles are formed in the arcing %%@
region and are then embedded in the growing film. The amount of amorphous overlay %%@
depends on the time they arrive with respect to the end of deposition. For an individual %%@
nanoparticle or a agglomerate of nanoparticles, the {\it aspect ratio} is the length (L) %%@
to width (D) ratio for the part protruding from the film surface. After chemical etching %%@
[see Fig.~\ref{etching}~(b)], the length of the protruding part might increase slightly, %%@
but the most prominent change is the decrease in its width. Both changes contribute %%@
positively to an increase in the aspect ratio. For a given set of deposition conditions, %%@
an optimum etching time is expected to exist (10 min in this case), for which the %%@
average amorphous overlay is removed. Obviously, etching for shorter periods will result %%@
in a partial removal of the amorphous overlayer and in a smaller number density of the %%@
exposed emission sites. On the other hand, chemical etching for prolonged periods will %%@
result in an excessive removal of amorphous material from the top and around the %%@
nanoparticles. Excessive loss of material from around the agglomerates is expected to %%@
result in them dropping out, thus decreasing the number density of emitting sites.
\section{Conclusions}
~~~ Carbon nitride thin films were deposited using a modified carbon arc method where a %%@
jet of N$_2$ gas was used to create a high pressure region in the vicinity of the %%@
electric arc. HREM analysis demonstrated that the material includes clusters of %%@
fullerene-like nanoparticles. Field emission measurements revealed that the resultant %%@
material has a relatively low threshold field and a high field emission current. %%@
Correlating the HREM and field emission results suggests that the good field emission %%@
properties are related to the protrusion of clusters of nanoparticles, which are %%@
topographically and electronically suitable for causing a concentration of the external %%@
electric field over their tips. This is further supported by the fact that the field %%@
emission characteristics of the nano-composite films examined were broadly similar to %%@
those achieved with nanotube suspension films. Chemical etching of the material %%@
deposited by the AJCA method resulted in a further improvement of the field emission %%@
characteristics. It is proposed that the removal of the amorphous material from the %%@
vicinity of the carbon nanoparticles during chemical etching results in an increase in %%@
the number of the emission sites and an effective enhancement of their aspect ratio. %%@
This consequently leads to easier emission of electrons from the protruding %%@
nanoparticles. These results suggest that the anodic jet carbon arc method can be used %%@
to deposit carbon nano-composite thin films which have field emission characteristics %%@
that are very similar to those obtained from pure carbon nanotubes.
\newpage
\bibliographystyle{prsty2}
\bibliography{bib}
\newpage
\newpage
\newpage
\newpage

\newpage

\newpage
\newpage
\newpage

\end{document}